\documentclass[twocolumn]{biophys}
\usepackage{amssymb,amsfonts,amsthm,bm}
\usepackage[round,numbers,sort&compress]{natbib}
\usepackage[ansinew]{inputenc}
\usepackage{comment}
\usepackage{multirow}
\usepackage{caption}
\usepackage{subcaption}
\usepackage{longtable}

\jno{kxl014}
\gridframe{N}
\cropmark{Y}
\doi{}%doi: 10.1529/biophysj.106.089789}
\newif\ifNOSUP \NOSUPfalse

\newif\ifNOFIG \NOFIGfalse

\newif\ifPAGEFIG \PAGEFIGfalse

\newif\ifPDFFIG \PDFFIGfalse

\def\thefigure{\arabic{figure}}

\begin{document}

%%%%%%%%%%%%%%%%%%%%%%%%%%%%%%%%%%%%%%%%%%%%%%%%%%
\title{Characterizing the folding core of the cyclophilin A --- cyclosporin A complex I: hydrogen exchange data and rigidity analysis}

\author{J. W. Heal$^{\ast}$$^{\dagger}$,
R. A. R\"{o}mer$^{\S}$, C. A. Blindauer$^{\ddagger}$ and R. B. Freedman$^{\P}$}
\address{$^{\dagger}$MOAC Doctoral Training Centre and Institute of Advanced Study, $^{\ddagger}$Department of Chemistry, $^{\S}$Department of Physics and Centre for Scientific Computing, and
  $^{\P}$School of Life Sciences, University of Warwick, Coventry, CV4 7AL,
  UK}% CV4 7AL}

%%%%%%%%%%%%%%%%%%%%%%%%%%%%%%%%%%%%%%%%%%%%%%%%%%
\begin{abstract}
{
The determination of a ``folding core'' can help to provide insight into the structure, flexibility, mobility and dynamics, and hence, ultimately, function of a protein --- a central concern of structural biology. Changes in the folding core upon ligand binding are of particular interest because they may be relevant to drug-induced functional changes. Cyclophilin A is a multi-functional ligand-binding protein and a significant drug target.
It acts principally as an enzyme during protein folding, but also as the primary binding partner for the immunosuppressant drug cyclosporin A (CsA).
Here, we have used hydrogen-deuterium exchange (HDX) NMR spectroscopy to determine the folding core of the CypA-CsA complex.
We also use the rapid computational tool of rigidity analysis, implemented in {\sc First}, to determine a theoretical folding core of the complex.
In addition we generate a theoretical folding core for the unbound protein and compare this with previously published HDX data.
The {\sc First} method gives a good prediction of the HDX folding core, but we find that it is not yet sufficiently sensitive to predict the effects of ligand binding on CypA.}
{$Revision: 1.38 $, compiled \today}%{Submitted February 22, 2008, and accepted for publication May 15, 2008.}%2
{*Correspondence: jack.heal@bristol.ac.uk \\
Address reprint requests to Jack Heal, University of Bristol, School of Chemistry, Cantock's Close, Bristol, BS8 1TS, UK.} 
%{*Correspondence: jack.heal@warwick.ac.uk \\
%Address reprint requests to Jack Heal, MOAC DTC, Senate House, University of Warwick, Coventry, CV4 7AL, UK.} %\\
%\\
%Editor:
%Richard W Aldrich.}%3
\end{abstract}
%%%%%%%%%%%%%%%%%%%%%%%%%%%%%%%%%%%%%%%%%%%%%%%%%%

\maketitle

\markboth{Heal et al.}{The CypA-CsA folding core I}

%%%%%%%%%%%%%%%%%%%%%%%%%%%%%%%%%%%%%%%%%%%%%%%%%%
\section*{INTRODUCTION}
%%%%%%%%%%%%%%%%%%%%%%%%%%%%%%%%%%%%%%%%%%%%%%%%%%

%%%%%%%%%%%%%%%%%%%%%%%%%%%%%%%%%%%%%%%%%%%%%%%%%%
%\section*{Folding cores}

The protein folding problem has
been a prevalent question during the past $50$ years as the emerging protein structure crucially determines flexibility, mobility and, ultimately, function \citep{DilM13}. The two principal competing theories on how protein
folding initiates are diffusion-collision \citep{KarW94} and
nucleation-condensation \citep{ItzOF95}. Indeed, it may well be that
both are valid depending on which protein is being investigated
\citep{HesRTK02}. Intuitively, residues which ``collapse early during folding'' \citep{Woo93}, might be particularly important to the overall folding process and are usually referred to as defining a \emph{folding core}. However, this set of residues is difficult to ascertain precisely.
One way of defining a
folding core experimentally is through $\Phi$-value analysis
\citep{FerMS92}. This approach focuses on the folding process by using point
mutations to determine the impact of particular residues on the energy
of the transition state in a one-step folding process. An alternative
 is to study the dynamics of the
folded structure through hydrogen-deuterium exchange (HDX) NMR
 experiments. For the
examples of barnase and chymotrypsin inhibitor 2, it has been shown
that the two definitions are consistent in that
slowly exchanging residues in HDX have high $\Phi$-values
\citep{LiW99}.

\begin{figure}[tbp] \begin{center}
    \includegraphics[width=\columnwidth]{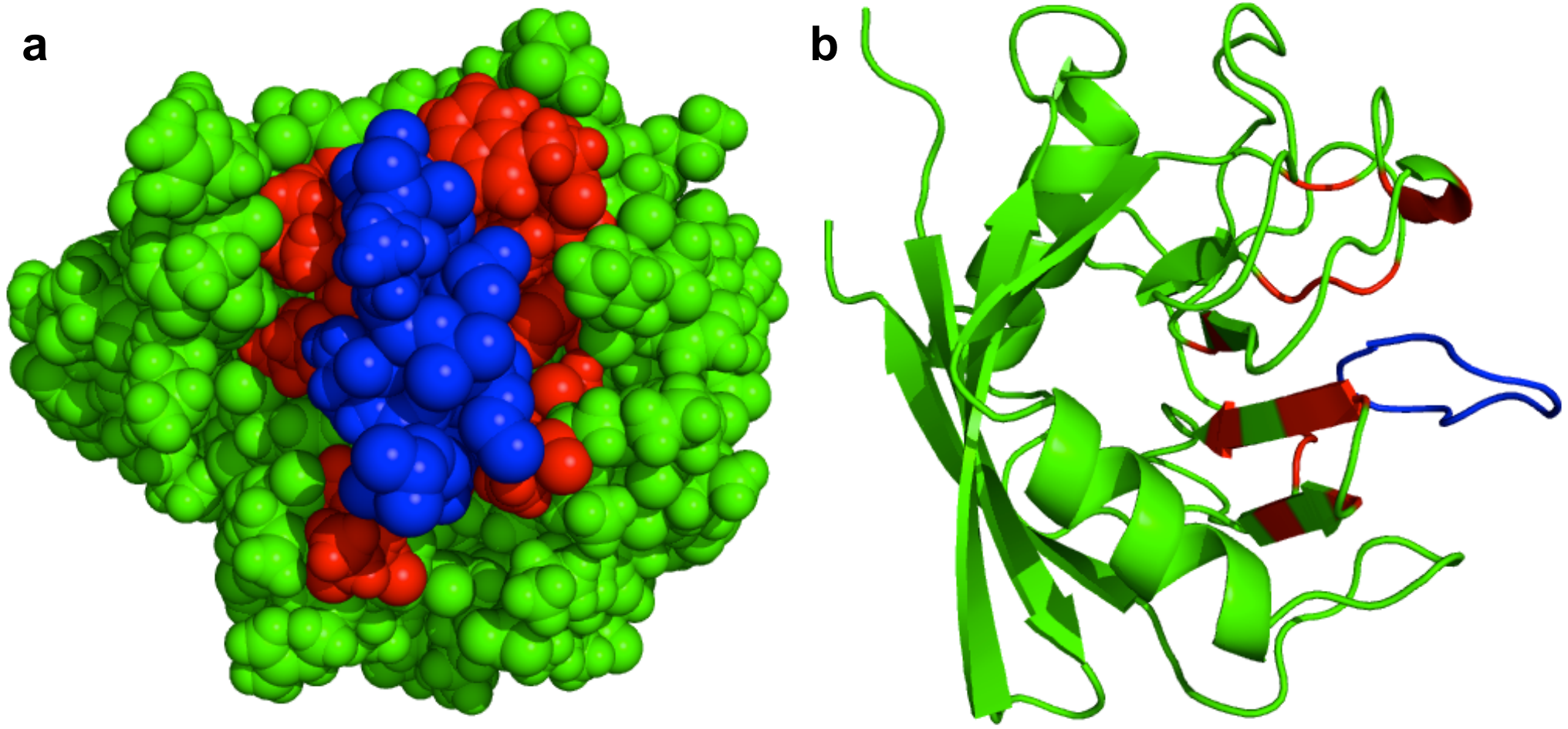}
    \caption[The structure of the CypA-CsA complex and the CypA binding site.]{(a) Sphere and (b) cartoon representation of the CypA-CsA complex  and the CypA binding site (using PDB structure 1CWA and the {\sc
        PyMOL} visualizer \citep{web_pymol}). CsA is indicated in blue and CypA in green. The 15 residues of CypA which have a heavy
      (non-hydrogen) atom within 4 \AA \, of the atoms in CsA are
      colored red.}
\label{fig:Binding_site}
\end{center}
\end{figure}

We have selected cyclophilin A (CypA), a multifunctional 18 kDa
protein with 165 residues, as the basis protein for our study since it is large enough to exhibit complex folding behaviour while at the same time it is readily investigated by HDX. Furthermore, among the large class of ligand-binding proteins, it is known to bind strongly to the immunosuppressant drug cyclosporin A (CsA)
\citep{NerMGO91,SpiBWW94,OttZGW97}, with dissociation constant $K_\mathrm{D} = 46$~nM \citep{LiuACS90}.
CypA acts as a peptidyl-prolyl {\it
  cis-trans} isomerase in addition to performing other roles when
binding to different molecules such as the HIV-1 capsid protein
\citep{LubBFK93,BosK04,WanH05}. The structure of the CypA-CsA
complex is shown in Figure~\ref{fig:Binding_site} with the binding site residues highlighted \citep{RadB11}.
%%%%%%%%%%%%%%%%%%%%%%%%%%%%%%%%%%%%%%%%%%%%%%%%%%
%%%%%%%%%%%%%%%%%%%%%%%%%%%%%%%%%%%%%%%%%%%%%%%%%%
Most commonly used to
suppress organ rejection following a transplant, CsA has also been
administered to treat ulcerative colitis, cardiac disease and a number
of autoimmune diseases \citep{NusP86,LicPKG94,MotZFM04}.
It is the CypA-CsA complex which binds to and inhibits the T-cell
activator calcineurin (CN) and thus has an immunosuppressant effect
\citep{ZydECF92,LiuJRR06}.

Here, we experimentally study the HDX behaviour of unbound CypA and also its complex with CsA. Using these HDX data, we establish the resulting folding cores. Our results compare very well with previously published HDX data on unbound CypA  \citep{ShiLHS06} and also elucidate the effect of ligand binding in the CypA-CsA complex.
NMR has been used to solve the structure of CypA \citep{OttZGW97} as
well as the CypA-CsA complex \citep{NerMGO91,SpiBWW94}. HDX
experiments on unbound CypA have previously been conducted
\citep{ShiLHS06} but the CypA-CsA complex has not previously been
studied in this way.

Establishing folding cores through $\Phi$-value
analysis or HDX provides valuable insight into protein folding and
dynamics, but also involves extensive experimental work.
For this reason, the prediction of HDX folding cores through rapid
computational methods is of ongoing interest
\citep{HesRTK02,RadB04,TarCV07,LiuPLH12,LobSDS13,ThoTEY13}. One method for predicting HDX folding cores uses rigidity analysis and is implemented in the {\sc First} software package \citep{HesRTK02,RadB04}. Having found the HDX folding core in our experiments, we next apply  {\sc First}
to unbound CypA and to the CypA-CsA complex.
We compare the resulting predicted folding cores with the HDX results.

%%%%%%%%%%%%%%%%%%%%%%%%%%%%%%%%%%%%%%%%%%%%%%%%%%
\section*{MATERIALS AND METHODS}
%%%%%%%%%%%%%%%%%%%%%%%%%%%%%%%%%%%%%%%%%%%%%%%%%%

%%%%%%%%%%%%%%%%%%%%%%%%%%%%%%%%%%%%%%%%%%%%%%%%%%
\section*{Protein expression and purification}

A derivative of the pQE-70 plasmid encoding for human CypA %expression was
was expressed in {\it E. coli} (JM109, New
England BioLabs) grown at 37$^{\circ}$C in minimal medium containing 1~g/L
($^{15}$NH$_4$)$_2$SO$_4$ (Cambridge Isotope Laboratories). When the
growth cultures reached an optical density of 0.5 at 600~nm, protein
expression was induced with 1 mM
isopropyl-$\beta$-D-thiogalactopyranoside (IPTG). Cells were harvested
after overnight growth during which selection pressure was maintained
by adding 0.1~mg/mL ampicilin.
Cell pellets were resuspended in 20 mM
4-(2-hydroxyethyl)-1-piperazineethanesulfonic acid (HEPES)
\label{HEPES} buffer at pH~$= 6.5$ and then stored at $-20^{\circ}$C.
Frozen cells were thawed, sonicated and then centrifuged, after which the
resulting supernatant was loaded onto a 10 mL Source 30S column (from
GE Healthcare) for cation exchange using 20 mM HEPES buffer at pH~$=
6.5$ and a concentration gradient of $0 - 150$ mM NaCl. Eluted
fractions of CypA were dialysed overnight against 10 mM ammonium
acetate buffer at pH~$= 6.5$ before being concentrated and
lyophilised.

%%%%%%%%%%%%%%%%%%%%%%%%%%%%%%%%%%%%%%%%%%%%%%%%%%
\section*{NMR assignment and HDX experiments}

For sequential assignment, lyophilised protein was resuspended in NMR buffer containing $4.2$ mM NaH$_2$PO$_4$, $15.8$ mM Na$_2$HPO$_4$ and $150$ mM NaCl at pH $6.5$. 2D [$^1$H,$^{15}$N] HSQC and 3D [$^1$H, $^{15}$N, $^1$H] TOCSY-HSQC and NOESY-HSQC data were acquired on a Bruker AV III 600 spectrometer operating at $600.13$ MHz for $^1$H and 60.81 MHz for $^{15}$N. 2D data were acquired with $16$ scans, $2048$ datapoints in F2 and $128$ increments in F1, and Fourier transformed with $2048\times 512$ datapoints over spectral widths of $16$ ppm in the $^1$H dimension (F2) and $42$ ppm in the $^{15}$N dimension (F1). 3D data were acquired with $8$ or $16$ scans and $2048 \times 40 \times 160$ datapoints in F3, F2 and F1, respectively, and transformed with $2048 \times 64 \times 512$ datapoints. Spectral widths were $16$ ppm in the $^1$H dimensions (F3, F1), and $38$ ppm in the $^{15}$N dimension (F2).
2D HSQC data for the HDX experiments were acquired on a Bruker AV II 700 spectrometer, equipped with a TCI cryoprobe, operating at $700.24$ MHz for $^1$H and  $70.96$ for $^{15}$N. The lyophilised protein was resuspended in NMR buffer as above, but made up in $99.9\%$ D$_2$O (Sigma-Aldrich). For the CypA-CsA sample, each spectrum was acquired with four scans, $2048$ points in F2, and $64$
increments in F1, and Fourier transformed with $2048 \times 256$ datapoints. The spectral widths were $16$ ppm and $42$ ppm in the $^1$H and $^{15}$N dimensions,
respectively. In all cases, chemical shifts $\delta$ were referenced to the residual HDO peak \citep{WisBYA95}.
Data were acquired  and processed using {\sc Topspin} version 2.1 (Bruker) and analyzed using {\sc Sparky} version 3.1 \citep{GodK}. 

%%%%%%%%%%%%%%%%%%%%%%%%%%%%%%%%%%%%%%%%%%%%%%%%%%
\section*{Rigidity analysis with {\sc First}}

Protein rigidity analysis is a computational method which rapidly
identifies rigid and flexible regions in a protein crystal structure
\citep{RadHKT01,JacRKT01}.
The structure is considered as a molecular framework in which bond
lengths and angles are considered fixed while dihedral angles are
permitted to vary. Atomic degrees of freedom are then matched against
bonding constraints
\citep{JacT95,JacH97,Jacobs98,JacKT99,ThoHYK00,JacRKT01,HesJT04}.
Covalent bonds, polar interactions (including hydrogen bonds and salt
bridges), and hydrophobic tethers can all be included as bonding
constraints. The output of the algorithm is a division of the
structure into rigid clusters and flexible regions, known as a \emph{rigid
cluster decomposition} (RCD). The RCD clearly depends on the
constraints present in the bond network, the strength and location of which are determined solely from the
geometry of the input structure.
A systematic removal of hydrogen bonds in order from weakest to strongest leads to a
loss in rigidity which we can relate to the unfolding of the protein \citep{RadHKT01}. This
is referred to as a \emph{rigidity dilution} (RD). The pattern of rigidity
loss can be used to gain insight into structural and functional
properties of the protein \citep{RadHKT01,WelJR09,HeaJWF12}. Indeed,
RDs have previously been used to predict the HDX folding core for a
number of proteins \citep{HesRTK02,RadB04} not including CypA.

The X-ray crystal structure 1CWA of the CypA-CsA complex was
downloaded from the Protein Data Bank (PDB) \cite{web_pdb}. Crystal
water molecules were removed before the {\sc reduce} software
\citep{WorLRR99} was used to add the hydrogen atoms and to flip side
chains of Asn, Gln and His residues where necessary. % to avoid steric
%clashes.
%
For simulations of the
unbound protein, CsA was deleted from the structure manually
using {\sc PyMOL} \cite{web_pymol}, which was used for
all molecular visualization.  
We note that the unbound structure for CypA is highly similar to the bound structure. Indeed, the backbone of 1CWA aligns with a different structure of the unbound protein (1W8V) with an RMSD of  0.26~\AA.

The strength of each hydrogen bond, measured in
kcal/mol, was calculated as a function of the geometry of the donor,
hydrogen and acceptor atoms using the highly distance- and angle-dependent Mayo potential \citep{DahGM97,WelJR09}.
Only hydrogen bonds with a bond
energy more negative than the energy cutoff parameter $E_{\mathrm{cut}}$ are included in the
network.
During an RD, $E_{\mathrm{cut}}$ is lowered causing some hydrogen bonds to be excluded
 from the network.
Rigidity dilution involves systematically lowering
$E_{\mathrm{cut}}$ and re-evaluating the RCD. Rigidity analysis was conducted using
{\sc First} version 6.1 \citep{web_flexweb}.

%%%%%%%%%%%%%%%%%%%%%%%%%%%%%%%%%%%%%%%%%%%%%%%%%%
\section*{RESULTS AND DISCUSSION}
%%%%%%%%%%%%%%%%%%%%%%%%%%%%%%%%%%%%%%%%%%%%%%%%%%

%%%%%%%%%%%%%%%%%%%%%%%%%%%%%%%%%%%%%%%%%%%%%%%%%%
\section*{Characterization of CypA and sequential assignment}

Purified recombinant CypA was characterized using mass spectrometry and circular
dichroism spectroscopy, confirming that the protein had the correct mass and secondary structure (Supporting Material).
When CypA binds to CsA, the change in environment of the Trp121
residue results in enhanced fluorescence (with an excitation wavelength of 290
nm and a peak of emission at 340 nm) \citep{HusZ94,GasVE99}. We
observed enhanced fluorescence during a titration of CsA into CypA,
which plateaued at a 1:1 concentration ratio (Supporting Material). This confirmed ligand binding.

Employing 3D TOCSY- and NOESY-HSQC data, and with the aid of previously published assignments for the CypA-CsA complex under different conditions \citep{NerMGO91,OttZGW97}, we have assigned 147 of the 159 non-proline residues of the protein in its unbound state as well as in  complex with CsA.

\begin{figure*}[tbh]
\begin{center}
  \includegraphics[width=\textwidth]{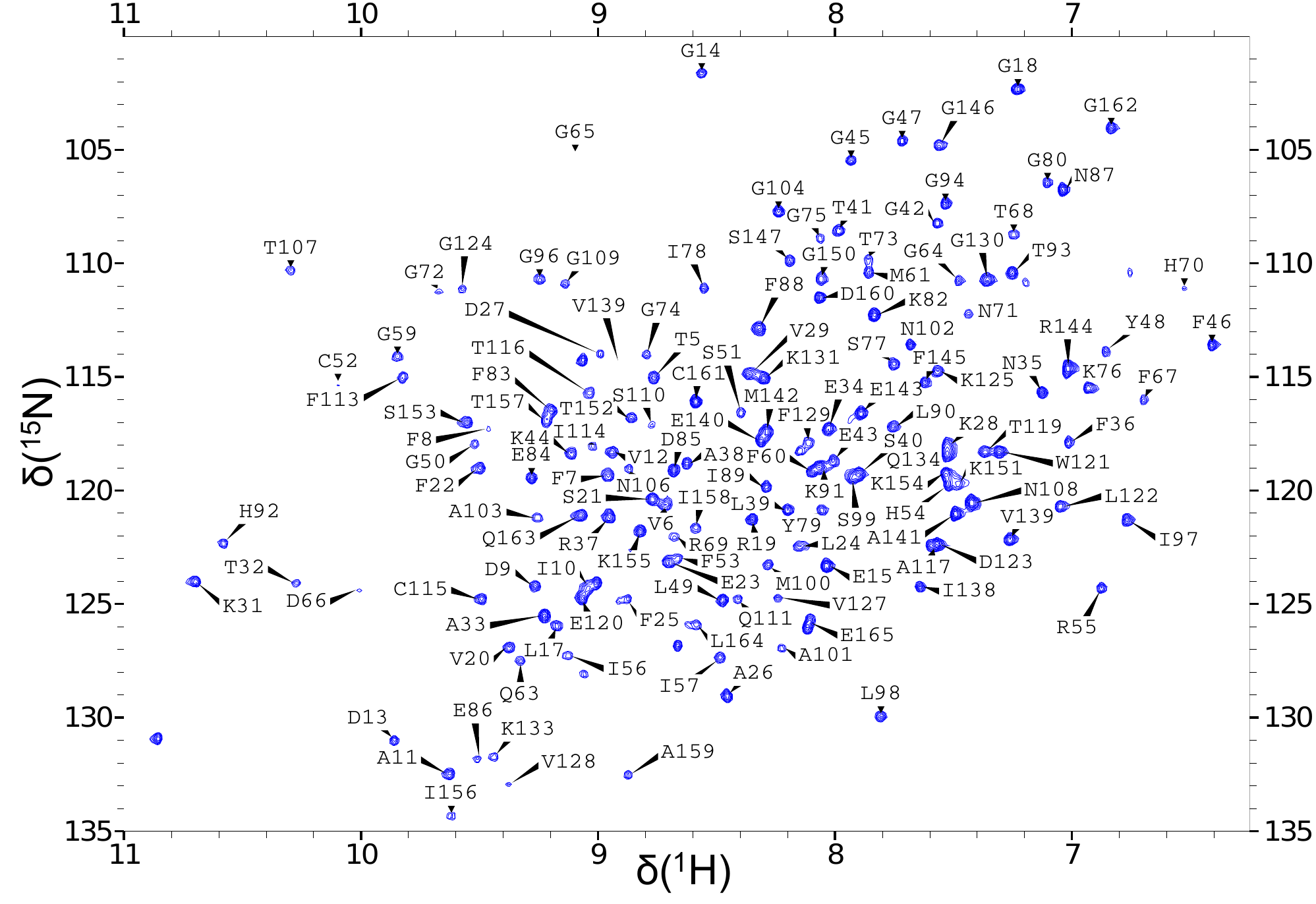}
  \caption[HSQC spectrum for the CypA-CsA complex]{HSQC spectrum of the
    CypA-CsA complex. Blue contour lines show signal intensity, and assigned backbone N-H cross peaks are labeled. Chemical shift frequencies $\delta$($^1$H) and $\delta$($^{15}$N) are given in parts per million (ppm).}
% Must take out spectra numbers. %With reference to Table \ref{tab:HDX_time}, spectra (b), (c) and (d) are spectra numbers 1, 15 and 32 respectively. }
\label{fig:HSQC}
\end{center}
%\end{sidewaysfigure}
\end{figure*}

%Backbone assignments of the CypA-CsA complex have been determined
%previously under different conditions \citep{NerMGO91,OttZGW97}. We
%have independently assigned 151 of the 159 non-proline residues of the protein in its unbound state as well as in  complex with CsA. 
The unassigned residues include residues $1$ -- $4$, which make up the flexible
N-terminus. In Figure \ref{fig:HSQC}, we show the assigned HSQC spectrum for the
CypA-CsA complex. Full lists of backbone N-H assignments for this spectrum and for the unbound protein
can be found in the Supporting Material.
%%%%%%%%%%%%%%%%%%%%%%%%%%%%%%%%%%%%%%%%%%%%%%%%%%
%%%%%%%%%%%%%%%%%%%%%%%%%%%%%%%%%%%%%%%%%%%%%%%%%%

%%%%%%%%%%%%%%%%%%%%%%%%%%%%%%%%%%%%%%%%%%%%%%%%%%
\section*{The HDX folding cores}

\begin{figure*}[tbh]
\begin{center}
  \includegraphics[width=\textwidth]{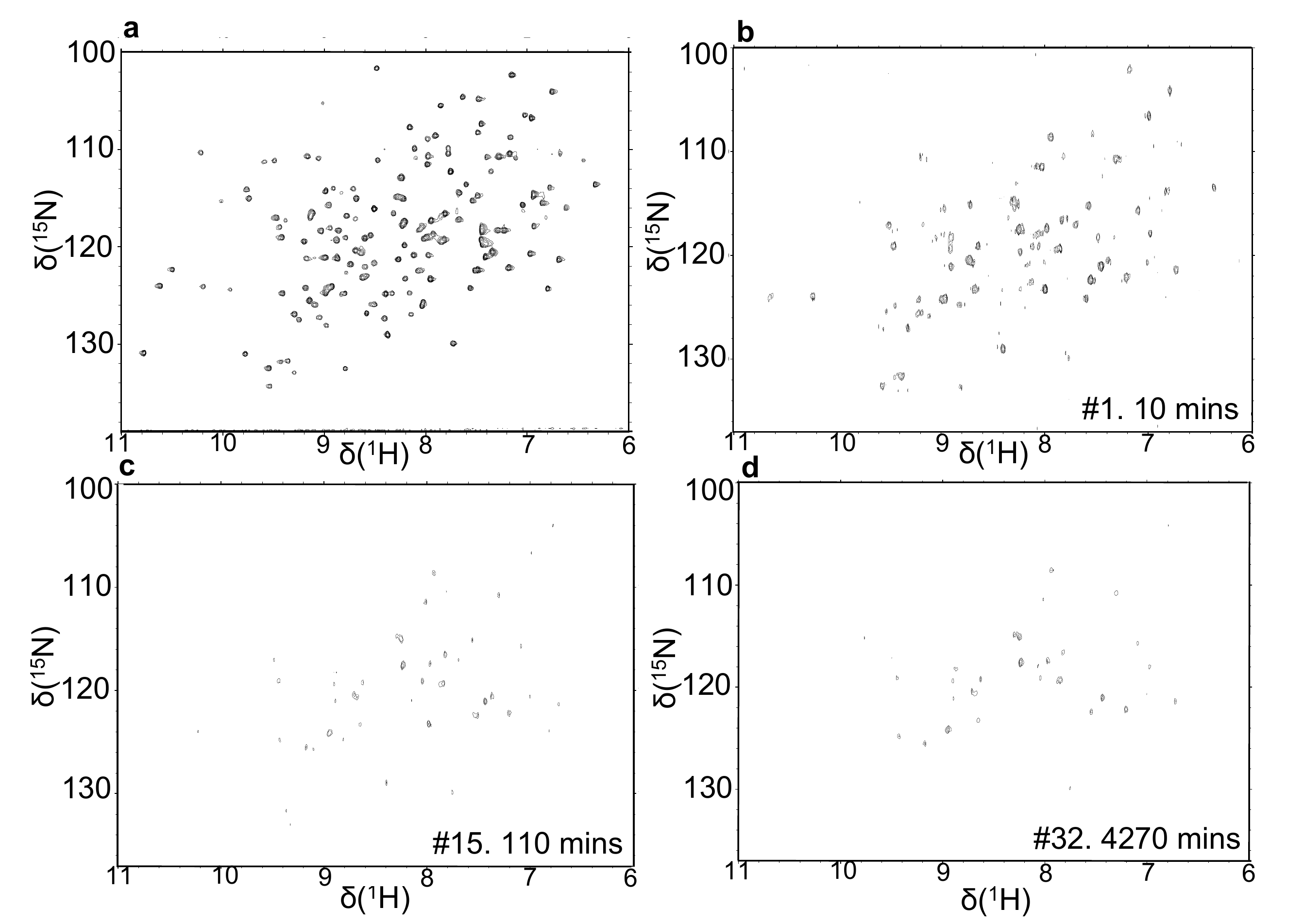}
  \caption[HSQC spectra of the CypA-CsA complex at different time
  intervals following initiation of HDX]%
{(a) HSQC spectrum of the
    CypA-CsA complex as in Figure \ref{fig:HSQC}. Following initiation of HDX, spectra were
    recorded after an elapsed time of, among others, (b) 10 minutes, (c) 110 minutes
    and (d) 4270
    minutes. In each spectrum, N-H cross peaks are shown as contour lines, representing signal intensity. Chemical shifts $\delta$($^1$H) and $\delta$($^{15}$N) are given in ppm.}
\label{fig:HDX_series}
\end{center}
%\end{sidewaysfigure}
\end{figure*}

\begin{figure*}[bt] \begin{center}
    \includegraphics[width=\textwidth]{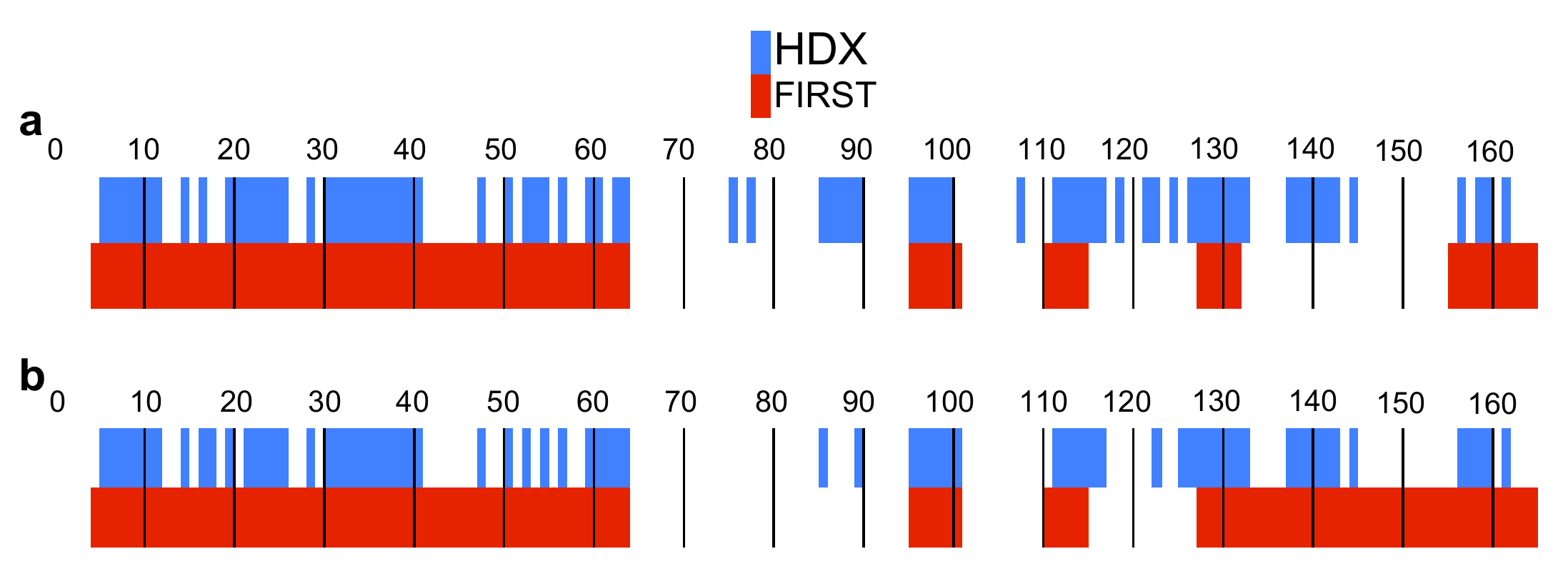}
%   \subfloat[Subcaption~1]{\label{figur:1}\includegraphics[width=\columnwidth]{../Figures/FIRST_HDX_proper.pdf}}
%  \subfloat[Subcaption 2]{\label{figur:2}\includegraphics[width=\columnwidth]{../Figures/FIRST_HDX_proper.pdf}}
    \caption[Comparison of folding cores]{The HDX folding cores (blue)
      are given along with the {\sc First} folding cores (red) for
      (a) the CypA-CsA complex and (b) unbound CypA. The residue
      numbers along the protein backbone are indicated, with thin vertical lines
      added every ten residues for clarity. Only the
      residues which are part of the folding cores are colored.}
\label{fig:FC_compare}
\end{center}
%\end{sidewaysfigure}
\end{figure*}

Figure~\ref{fig:HDX_series} shows the HSQC spectrum of the CypA-CsA
complex alongside spectra recorded 10, 110 and 4270 minutes (71 hours
and 10 minutes) after initiating HDX.
%%%%%%%%%%%%%%%%%%%%%%%%%%%%%%%%%%%%%%%%%%%%%%%%%%
%%%%%%%%%%%%%%%%%%%%%%%%%%%%%%%%%%%%%%%%%%%%%%%%%%
We define the set of residues for which a corresponding backbone amide signal remains in the HSQC
spectrum after $110$ minutes to be the experimentally determined HDX folding core of the complex. These
residues are listed in Table \ref{tab:HDX_FC} and also indicated in  Figure~\ref{fig:FC_compare} (a) and (b).

The published HDX experiments on unbound CypA resulted in a
classification of CypA residues in terms of their exchange rates,
$k_{\mathrm{ex}}$ \citep{ShiLHS06}.
Twelve residues, including the proline residues, were not
categorized since they were not identified in the HSQC spectrum. 
We also carried out HDX experiments on the unbound protein, and our data,  shown in the Supporting Material, were in agreement with the previously published result \citep{ShiLHS06}.

Here we have drawn on the exchange rates determined in \citep{ShiLHS06} and defined the residues with $k_{\mathrm{ex}} < 10^{-2}$ min$^{-1}$ as the folding core of the unbound protein, a definition congruent with that applied to our dataset for the CypA-CsA complex (see above).
According to this approach, the
HDX folding core for unbound CypA has $73$ residues.
%

%%%%%%%%%%%%%%%%%%%%%%%%%%%%%%
\begin{table}[bt] \centering \caption[The HDX folding core residues for
  the CypA-CsA complex]{The residues of the HDX folding core for the CypA-CsA
    complex. Underlined residues are slowly exchanging only in the presence of the ligand.}
 \label{tab:FC_both}
%\begin{center}
{\begin{tabular}{|p{8cm}|}
\hline
CypA-CsA HDX folding core residues \\
\hline\hline
 V6, 	F7, 	F8, 	D9, 	I10, {A11}, 	V12, 	E15, {L17}, V20, \emph{S21}, 	F22, 	 E23, 	L24, 	F25, 	{A26}, V29, {K31}, 	T32, 	A33, 	E34, 	N35, 	F36, 	R37, 	 A38, 	L39, 	S40, 	T41, 	{Y48}, 	S51, F53, {\emph{H54}}, {R55}, {I57}, 	F60, 	M61, 	 Q63, 	{G64}, {\emph{K76}}, {\emph{I78}}, {E86}, {\emph{N87}}, {\emph{F88}}, {\emph{I89}}, {L90}, {G96}, 	I97, 	L98, {S99}, {M100}, {\emph{N108}}, F112, {F113}, 	{I114}, C115, {T116}, 	A117, {\emph{T119}}, {\emph{L122}}, 	D123, {K125}, {V127}, {V128}, 	F129, 	 G130, 	K131, 	V132, 	K133,  {I138},	V139, 	E140, 	A141, 	M142, 	E143, 	F145, T157, {A159}, D160, {G162}  \\
\hline
\end{tabular}}{}
\label{tab:HDX_FC}
\end{table}
%%%%%%%%%%%%%%%%%%%%%%%%%%%%%%
%%%%%%%%%%%%%%%%%%%%%%%%%%%%%%%%%%%%%%%%%%%%%%%%%%
%%%%%%%%%%%%%%%%%%%%%%%%%%%%%%%%%%%%%%%%%%%%%%%%%%
We find that the HDX folding core does not change dramatically upon ligand binding. Its size increases from $73$ to $80$ residues. This is due to the ten residues, highlighted in Table \ref{tab:HDX_FC}, which are slowly exchanging in the presence of the CsA ligand but not part of the HDX folding core for unbound CypA. However, there are also three assigned residues  --- G18, A101 and I158 --- which are part of the 'unbound' folding core of \citep{ShiLHS06} but are no longer part of the 'bound' folding core. This slight increase in folding core size is consistent with the expectation that the presence of a ligand is likely to shield certain residues from the surface of the protein.
Eight of the highlighted residues belong to flexible regions which are proximal to the binding site. Hence those changes are consistent with a restricted flexibility of the unstructured regions near the binding site in the presence of the ligand.
The other two residues in the set, S21 and H54, are located in $\beta$-strands.
We note that these residues are surrounded by folding core residues (V20, F22, F53, R55) and so it is somewhat unexpected that these are not part of the folding core for the unbound protein. That they vanish earlier than expected from the HDX experiments on the unbound protein may be due to the inherent low signal intensity in these experiments. 
%That these two residues become part of the folding core for the CypA-CsA complex is somewhat unexpected.

%%%%%%%%%%%%%%%%%%%%%%%%%%%%%%%%%%%%%%%%%%%%%%%%%%
\section*{The {\sc First} folding cores}

We then simulated RDs by systematically lowering $E_{\mathrm{cut}}$,
i.e. removing the hydrogen bonds in order of strength from weakest to strongest
\citep{WelJR09}. We used {\sc First} to generate an RCD each time a
hydrogen bond was removed. In a 1D representation of an RCD, each
residue in the primary structure is labelled as being rigid or
flexible depending on the rigidity of its C$_{\alpha}$ atom. We show
rigid residues as blocks which are colored according to their rigid
cluster membership.

\begin{figure}[hp]
\begin{center}
    \includegraphics[width=\columnwidth]{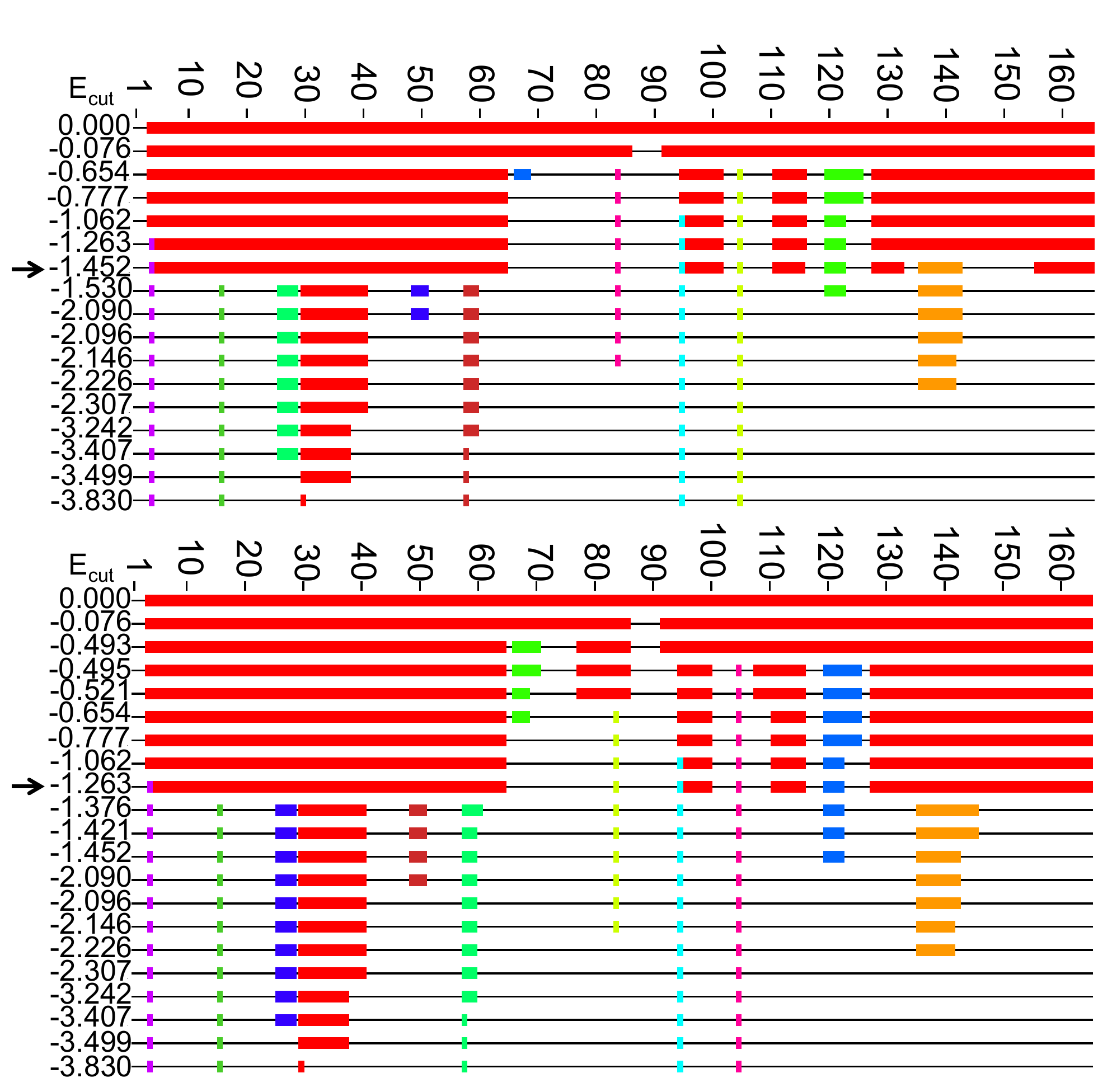}
    \caption[RD plots of CypA]{RD plots of (a) the
      CypA-CsA complex, and (b) unbound CypA. The RCD is shown at
      different values of $E_{\mathrm{cut}}$. The units of $E_{\mathrm{cut}}$ are kcal/mol. Rigid
      residues are shown as thick colored blocks and flexible regions as
      thin horizontal black lines. Residues which are mutually rigid are shown in
      the same color. The line representing the {\sc First} folding core in each case
       is indicated with an arrow.}
\label{fig:CypA_RCDs}
\end{center}
\end{figure}

We visualize the pattern of rigidity loss during RDs by plotting the
1D representation of the RCD each time this changes. Such plots for
the CypA-CsA complex and the unbound CypA are given in
Figure~\ref{fig:CypA_RCDs}. When $|E_{\mathrm{cut}}|$ is small, the
protein is largely rigid and many of the residues are represented as
blocks. As $E_{\mathrm{cut}}$ becomes more negative, i.e.\ as stronger bonds are excluded from the bond network, more residues
become flexible.
%%%%%%%%%%%%%%%%%%%%%%%%%%%%%%%%%%%%%%%%%%%%%%%%%%%
%%%%%%%%%%%%%%%%%%%%%%%%%%%%%%%%%%%%%%%%%%%%%%%%%%%
%
In both of the RD plots there is a clear and abrupt transition from
the largely rigid state to the largely flexible state, consistent with
the first-order rigidity loss expected for a predominantly
$\beta$-sheet protein \citep{WelJR09}. The lowest line in the RD plot
where at least three residues of two or more secondary structures (as
determined using the DSSP algorithm \citep{KabS83}) are
part of the same rigid cluster determines the {\sc First} folding core
\citep{HesRTK02,RadB04,RadAIK04,TasYGA07}. %The secondary structure was
%determined using the DSSP algorithm \citep{KabS83}.
We refer to the
$E_{\mathrm{cut}}$ corresponding to this line as the folding core
energy, $E_{\mathrm{fc}}$. For unbound CypA, $E_{\mathrm{fc}}$ =
$-1.263$ kcal/mol and for the CypA-CsA complex, $E_{\mathrm{fc}}$ =
$-1.452$ kcal/mol. That the CypA-CsA complex has a lower
$E_{\mathrm{fc}}$ suggests that ligand binding confers stability on
the complex, as more bonds need to be broken in order to render the
protein mostly flexible \citep{TasYGA07}. The residues that are
mutually rigid in the RCD evaluated at $E_{\mathrm{fc}}$,
colored red in Figure~\ref{fig:CypA_RCDs},
  form the {\sc  First} folding core.

%%%%%%%%%%%%%%%%%%%%%%%%%%%%%%%%%%%%%%%%%%%%%%%%%%
\section*{Comparison of {\sc First} and HDX folding cores}

Figure~\ref{fig:FC_compare} allows us to compare the HDX and {\sc First} folding
cores along the primary structure of the CypA-CsA complex and unbound
CypA. Residues which form part of the folding cores are represented as
colored blocks. In both cases, the {\sc First} folding cores largely
overlap with the HDX folding cores.
The small changes which do
occur upon ligand binding in the HDX folding cores are not very well captured in the two {\sc First} folding cores. Rather, these
differ only between residues $133$ and $155$, whereas the HDX folding cores remain largely unaffected in this region.
Notably, the {\sc First} folding core for the
CypA-CsA complex is smaller than for the unbound protein. This contrasts with
our expectation and experimental finding as given above, where we show that the HDX folding core increases
in size upon ligand binding.
This highlights a problem with a theoretical method that only uses {\sc First}. Ligand binding to the surface of CypA causes the binding site residues to become buried where before they were exposed, which may affect their HDX exchange rates. This effect is not modelled in {\sc First}, where we merely consider the hydrogen bond network of the initial, static crystal structure. 

\begin{figure*}[tbh] \begin{center}
    \includegraphics[width=\textwidth]{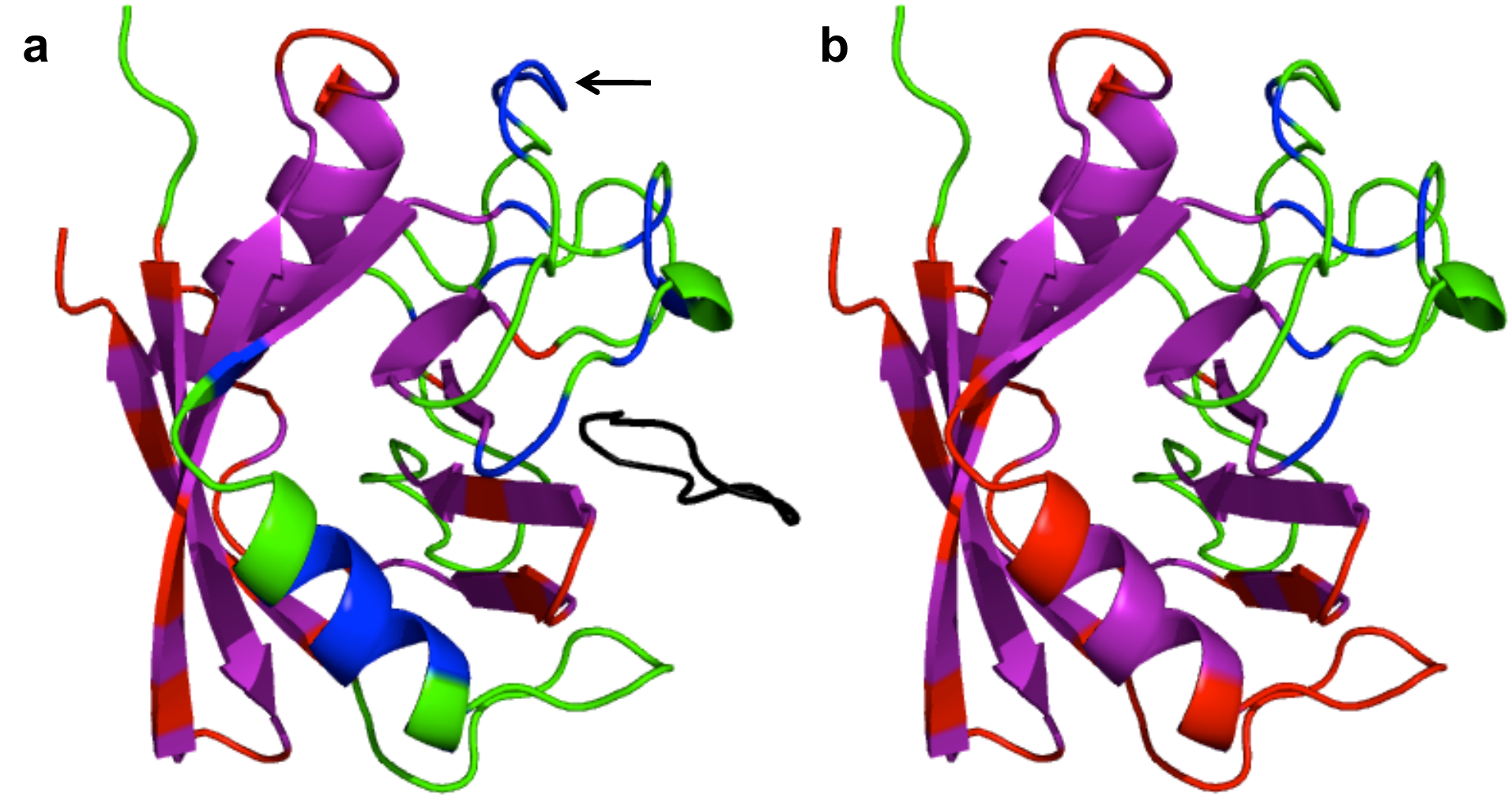}
    \caption[Comparison of folding cores in {\sc PyMOL}]{The HDX and {\sc
        First} folding cores for (a) the CypA-CsA complex and (b) unbound
      CypA indicated in the cartoon representation. Residues are colored purple when they are part of both folding cores,
      blue when they belong to the HDX folding core only, red when part of the {\sc First} folding core
      only or green  when not part of either folding core. In (a), CsA is
      colored black and the arrow indicates the loop region containing residues 87 -- 89.} \label{fig:1CWA_FCs} \end{center}
%\end{sidewaysfigure}
\end{figure*}

In Figure~\ref{fig:1CWA_FCs} the HDX and {\sc First} folding cores are shown superimposed onto the
structure 1CWA before and after the removal of the ligand. We see that despite the issues addressed above, theoretically and experimentally determined folding cores agree reasonably well  for most of the residues. %, as indicated by the large sections of the protein colored either green or purple.
The helix in the foreground of each image shows a marked difference when the ligand is removed, caused by the difference in {\sc First} folding core between residues 133 and 155. Residues 87 -- 89 form part of the loop indicated by the arrow in Figure~\ref{fig:1CWA_FCs}(a), and are only part of the HDX folding core in the presence of CsA.
%%%%%%%%%%%%%%%%%%%%%%%%%%%%%%%%%%%%%%%%%%%%%%%%%%
%%%%%%%%%%%%%%%%%%%%%%%%%%%%%%%%%%%%%%%%%%%%%%%%%%

%%%%%%%%%%%%%%%%%%%%%%%%%%%%%%%%%%%%%%%%%%%%%%%%%%
\section*{CONCLUSION}
%%%%%%%%%%%%%%%%%%%%%%%%%%%%%%%%%%%%%%%%%%%%%%%%%%

The experimental HDX folding core for the CypA-CsA complex is highly
similar to the folding core for the unbound protein, albeit with a small number of additional residues. %The number of residues in the HDX folding core for CypA increases when in complex with CsA.
This small change is consistent with previous observations of only subtle conformational change in CypA upon ligand binding \citep{FanF02,OttZGW97}. The {\sc First} folding cores differ more substantially. Ligand binding confers rigidity upon the structure but alters the pattern of rigidity loss so that the {\sc First} folding core in fact decreases in size.
In both cases, the {\sc First} folding core is a reasonable match for the HDX folding core,
in agreement with previous folding core predictions using {\sc First}
\citep{HesRTK02}. The {\sc First} folding core is defined by
 the RCD at $E_{\mathrm{fc}}$, a value which decreases upon ligand binding.
 This means that when the folding cores are compared
 for the protein before and after ligand removal, we are
comparing RCDs at different $E_{\mathrm{cut}}$. As a result there may
be more constraints present in the unbound protein $E_{\mathrm{cut}}$
than for the complex, resulting in the unlikely prediction of a larger
folding core in the absence of a ligand.
{\sc First} is a rapid tool to implement and this study complements
previous studies which demonstrate its utility for folding core
prediction \citep{HesRTK02,RadB04,RadAIK04,TasYGA07}. Nevertheless, our work also
shows that the {\sc First}-based folding core predictions are not yet accurate or
sensitive enough to capture the impact of ligand binding for CypA. For the purpose of
predicting subtle effects of ligand binding, just analyzing patterns of rigidity as done through {\sc First} appears insufficient.
The trade-off between, on one hand, rapid computation as achieved with {\sc First} %and simplicity of input information as in {\sc First}-based approach
 and, on the other hand, the necessary accuracy of folding core prediction needs to be more finely balanced. In the related Ref.\ \cite{HeaWFR14}, we show that when taking into account not only the rigidity of a protein, but also its propensity for motion, improved theoretical predictions for folding cores can be made with greater sensitivity and specificity. % --- albeit still not with perfect accuracy.

%Balancing computational complexity with accuracy \citep{TarCV07,LobSDS13}

%%%%%%%%%%%%%%%%%%%%%%%%%%%%%%%%%%%%%%%%%%%%%%%%%%%
%\section*{SUPPLEMENTARY MATERIAL}
%%%%%%%%%%%%%%%%%%%%%%%%%%%%%%%%%%%%%%%%%%%%%%%%%%%
%
%An online supplement to this article can be found by visiting BJ Online at http://www.biophysj.org.

%%%%%%%%%%%%%%%%%%%%%%%%%%%%%%%%%%%%%%%%%%%%%%%%%%
\section*{ACKNOWLEDGMENTS}

{\ack We thank S.A.\ Wells for help with the modeling aspects of this work. The CypA plasmid was kindly provided by G.\ Fischer from the Max Planck Research Unit for Enzymology of Protein Folding in Halle, Germany. We gratefully acknowledge funding from the EPSRC Life Sciences Interface programme (MOAC DTC EP/F500378/1). JWH thanks the Institute of Advanced Study for its support in the form of an Early Career Fellowship.}\vspace*{6pt}

%%%%%%%%%%%%%%%%%%%%%%%%%%%%%%%%%%%%%%%%%%%%%%%%%%%%%%%%%%%%
%\subsection*{Supporting Citations} References \cite{LiWJR12,AmiWWR13,LiwCOS08,HawDF91} appear in the Supporting Material.

%%%%%%%%%%%%%%%%%%%%%%%%%%%%%%%%%%%%%%%%%%%%%%%%%%
% Bibliography style (requires the style file biophysj.bst in the
% document directory)
\bibliographystyle{biophysj}\bibliography{bibliograph}

\ifPAGEFIG
\pagestyle{empty}
%%%%%%%%%%%%%%%%%%%%%%%%%%%%%%%%%%%%%%%%%%%%%%%%%%%%%%%%%%%%
% Figures
%%%%%%%%%%%%%%%%%%%%%%%%%%%%%%%%%%%%%%%%%%%%%%%%%%%%%%%%%%%%
\clearpage

\setcounter{figure}{0}

\newpage

\fi% PAGEFIG

% end here if supplement is not included
\ifNOSUP\end{document}\else%

%%%%%%%%%%%%%%%%%%%%%%%%%%%%%%%%%%%%%%%%%%%%%%%%%%%%%%%%%%%%
%%
%% Supporting Material
%%
%% to be moved to its own file later
%%
%%%%%%%%%%%%%%%%%%%%%%%%%%%%%%%%%%%%%%%%%%%%%%%%%%%%%%%%%%%%
\clearpage\newpage
\onecolumn
\setcounter{figure}{0}
\setcounter{table}{0}
\def\thefigure{S\arabic{figure}}
\def\thetable{S\arabic{table}}
\setcounter{page}{1}
%\pagestyle{plain}
%%%%%%%%%%%%%%%%%%%%%%%%%%%%%%%%%%%%%%%%%%%%%%%%%%%%%%%%%%%%

\section*{Supporting Material}

\textbf{Characterizing the folding core of the cyclophilin A --- cyclosporin A complex I: hydrogen exchange data and rigidity analysis}\\

\noindent%
J. W. Heal, R. A. R\"{o}mer, C. A. Blindauer and R. B. Freedman

%%%%%%%%%%%%%%%%%%%%%%%%%%%%%%%%%%%%%%%%%%%%%%%%%%%%%%%%%%%%
\section{Mass spectrometry and circular dichroism spectroscopy}

\begin{figure}[p]
\begin{center}
\includegraphics[width=\columnwidth]{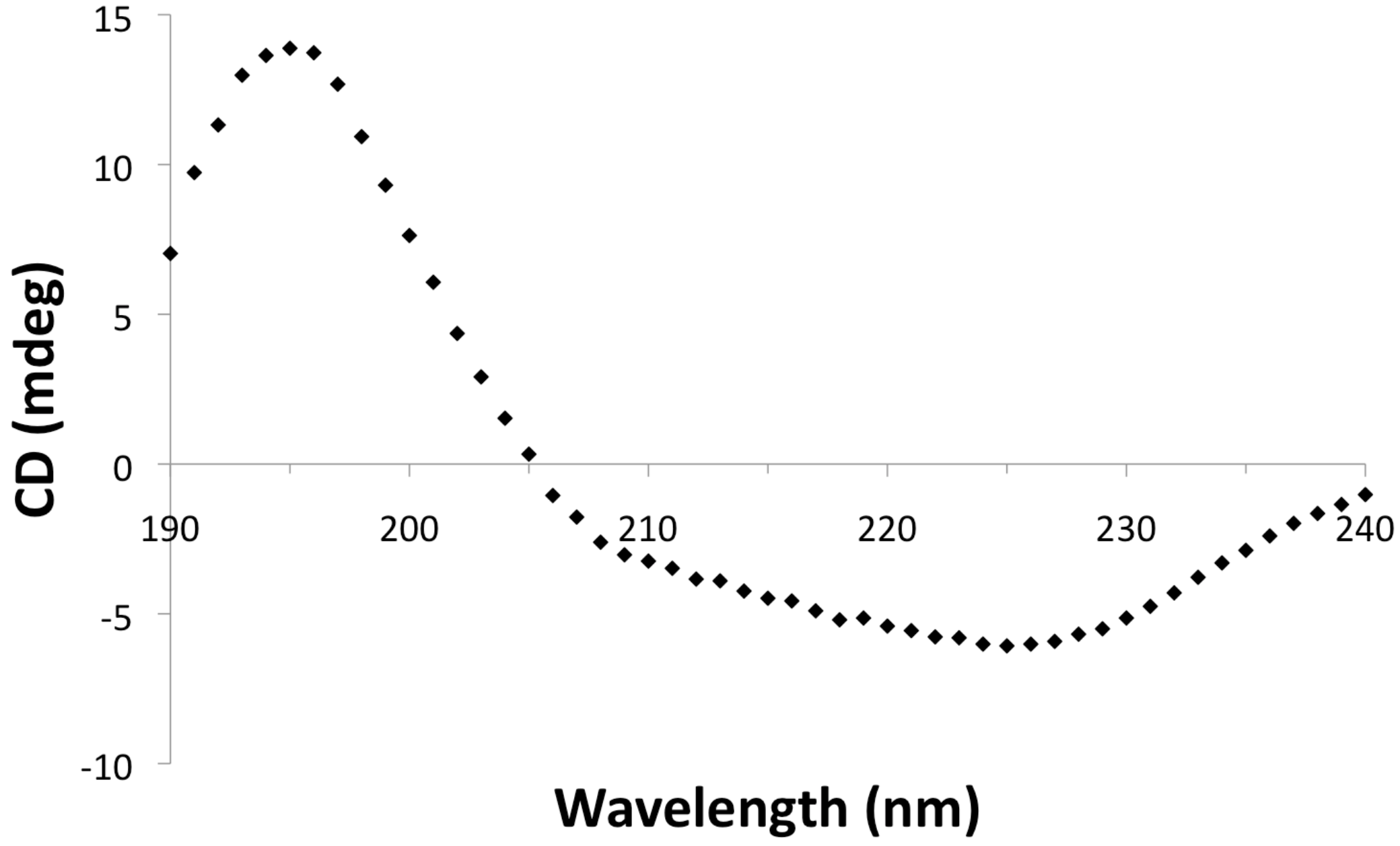}
\caption[The far-UV CD spectrum of CypA]{Far-UV CD spectrum of CypA at 25$^{\circ}$C. The average CD signal from 16 scans is plotted at 1~nm intervals between 190~nm and 240~nm.} %Data was processed using Dichroweb \citep{WhiW04}.}
\label{fig:CD_fold}
\end{center}
\end{figure}

To estimate the secondary structure composition for purified CypA in order to verify that it was correctly folded, the far-UV CD spectrum of CypA was measured at  25$^{\circ}$C using a Jasco J-815 CD spectropolarimeter. The resulting spectrum is shown in Figure \ref{fig:CD_fold}. 
%%%%%%%%%%%%%%%%%%%%%%%%%%%%%%%%%%%%%%%%%%%%%%%%%%
%%%%%%%%%%%%%%%%%%%%%%%%%%%%%%%%%%%%%%%%%%%%%%%%%%
%
CD spectra were recorded for protein samples of 0.1~mg/mL in 5~mM sodium phosphate, pH~7.3 unless otherwise stated. Small volumes of concentrated purified protein were diluted in sodium phosphate buffer and the resulting concentration confirmed by measuring $A_{280}$. 
Data points were collected during far-UV scans between 180~nm and 260~nm at 1~nm intervals.  A baseline spectrum was recorded in the same way for a sample containing buffer only.  To generate Figure \ref{fig:CD_fold}, 16 scans were collected at 100~nm/min in continuous scanning mode, and the average CD signal minus the average baseline  signal was plotted. We show the data collected between 190~nm and 240~nm at 1~nm intervals. 
The data was analysed with Dichroweb, using the reference database SP175 to calculate the proportional secondary structure composition of CypA. Our protein sample was determined to be 18~$\%$ helix and 33~$\%$ sheet. We compared this with the results of the DSSP algorithm applied to five X-ray crystal structures in the PDB. The average composition of these structures was 13~$\%$ $\alpha$-helix and 32~$\%$ $\beta$-sheet.

%%%%%%%%%%%%%%%%%%%%%%%%%%%%%%%%%%%%%%%%%%%%%%%%%%%%%%%%%%%%
\section{Fluorescence}

\begin{figure}[p]
\begin{center}
\includegraphics[width=0.9\columnwidth]{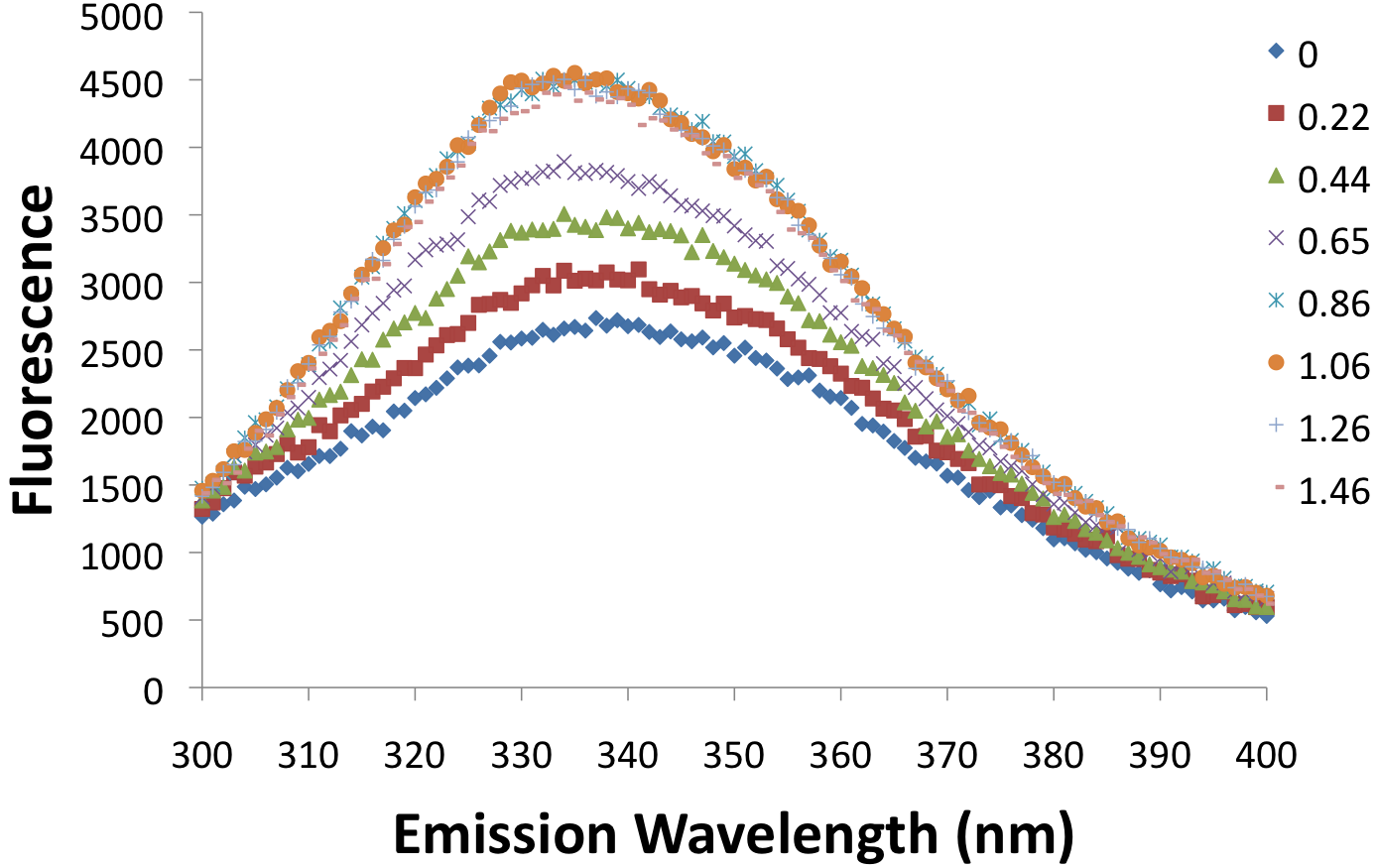}
\caption[Fluorescence emission spectra of CypA with increasing CsA concentration]{Fluorescence emission spectra of CypA with increasing CsA concentration. The key shows the concentration ratio [CsA]/[CypA]. Emission spectra were recorded at 1~nm intervals between 300~nm and 400~nm with excitation wavelength of 290~nm. The average emission value of three scans is plotted.}
\label{fig:Fluo_curves}
\end{center}
\end{figure}

When CsA binds to CypA, Trp121 becomes shielded from the solvent and its fluorescence increases as a result \citep{LiuACS90,HusZ94}. 
CsA was titrated into a solution of CypA and fluorescence spectroscopy was used to monitor the change in tryptophan fluorescence. 
The experiment was conducted using a Photon Technology International fluorimeter. A stock solution  of CypA (65 $\mu$M) was diluted 1~mL in 50~mM TRIS buffer at pH 7.3 so that the final CypA concentration was 4.5~$\mu$M. 
CsA was stored in ethanol at a concentration of 1.0~mM. A stock solution of 0.1~mM CsA for the titration was made by diluting this ten-fold in TRIS buffer. Each titre consisted of 10~$\mu$L of CsA stock, and  so increased the total concentration of CsA in the fluorescence sample by approximately 1~$\mu$M. 
After each increment in [CsA] the  sample was mixed using a pipette. The emission spectra were recorded at 1~nm intervals between 300~nm and 400~nm, using an excitation wavelength of 290 nm. The average fluorescence emission of three scans was recorded for each wavelength, using a slit width of 2~nm. Two baselines were recorded, for samples containing TRIS buffer only and 5~$\mu$M CsA in TRIS buffer. There was no significant difference between the baselines (data not shown). 
Baseline adjusted fluorescence emission spectra for CypA with various concentrations of CsA are shown in Figure \ref{fig:Fluo_curves}. The key shows the concentration fraction [CsA]/[CypA]. We observe a steady increase in fluorescence with increasing [CsA] until [CsA]/[CypA] = 1. After this point is reached, adding more CsA does not enhance fluorescence. 
%%%%%%%%%%%%%%%%%%%%%%%%%%%%%%%%%%%%%%%%%%%%%%%%%%
%%%%%%%%%%%%%%%%%%%%%%%%%%%%%%%%%%%%%%%%%%%%%%%%%%

%%%%%%%%%%%%%%%%%%%%%%%%%%%%%%%%%%%%%%%%%%%%%%%%%%%%%%%%%%%%
\section{HDX results for unbound CypA}

\begin{figure*}[p]
\begin{center}
\includegraphics[width=0.85\textwidth]{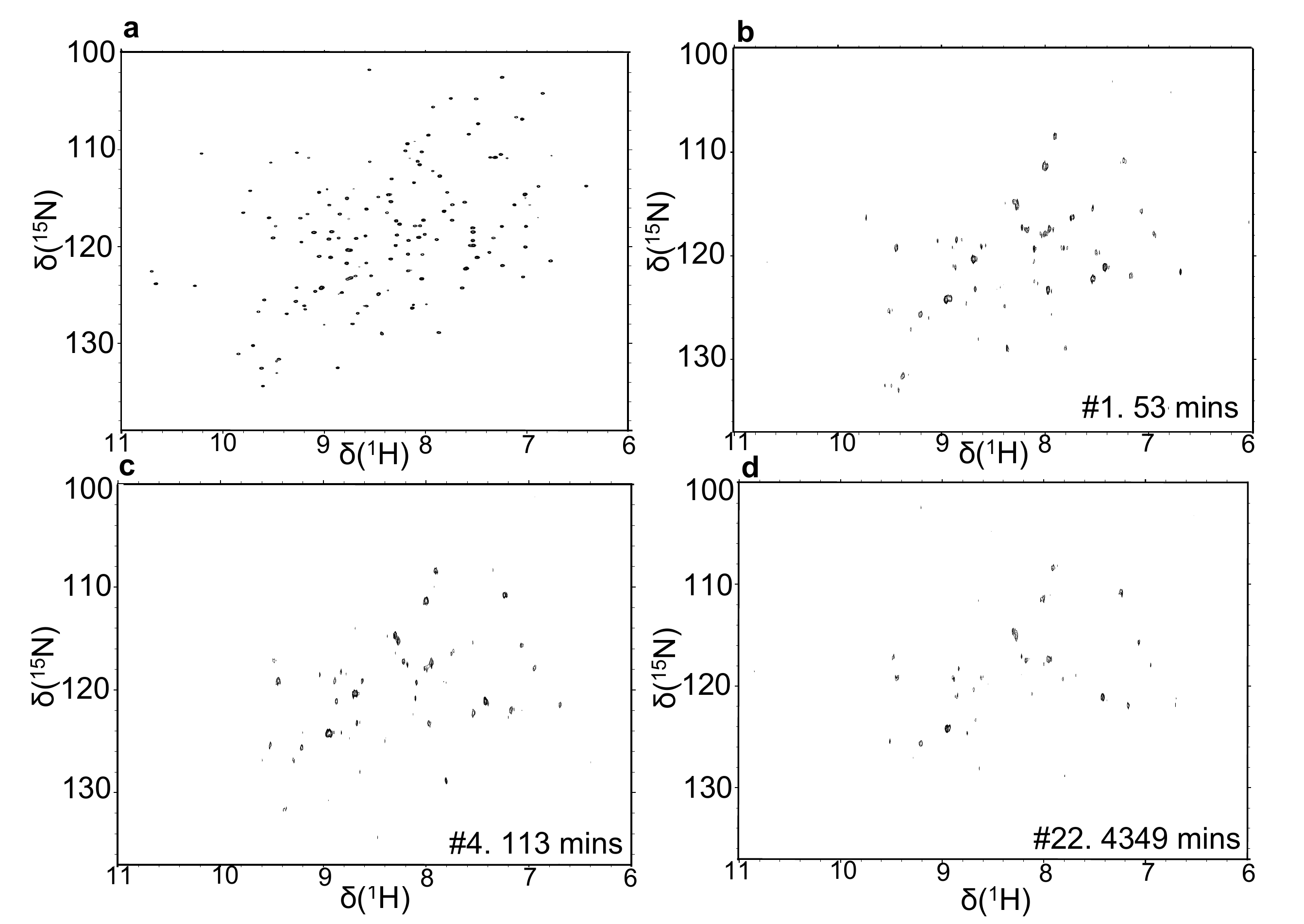}
\caption[HSQC spectra of unbound CypA at different time intervals following initiation of HDX]{(a) HSQC spectrum of unbound CypA. HSQC spectra are also shown at different time intervals following initiation of HDX. These are (b) 53 minutes, (c) 113 minutes and (d) 4349 minutes.}
\label{fig:HDX_noCsA}
\end{center}
\end{figure*}

For unbound CypA, we increased the number of scans recorded for each spectrum from 4 to 16 in response to a lower yield in order to maximise the signal to noise ratio. 
The earliest usable HSQC spectrum was completed 53 minutes after adding D$_2$O to the protein sample, and the subsequent spectra were therefore recorded less frequently than with the CypA-CsA experiments.  
%%%%%%%%%%%%%%%%%%%%%%%%%%%%%%
Figure \ref{fig:HDX_noCsA} shows the full HSQC spectrum, along with spectra taken at different time intervals following the initiation of HDX. Spectrum numbers 1, 4 and 22 are shown, recorded after 53, 113 and 4349 minutes respectively. 
%%%%%%%%%%%%%%%%%%%%%%%%%%%%%%%%%%%%%%%%%%%%%%%%%%
%%%%%%%%%%%%%%%%%%%%%%%%%%%%%%%%%%%%%%%%%%%%%%%%%%

\section{N-H assignment tables}

Table \ref{tab:assign} gives the chemical shift assignments for each residue of unbound protein as well as the CypA-CsA complex.

%%%%%%%%%%%%%%%%%%%%%%%%%%%%%%%%%%%%%%%%%%%%%%%%%%
\begin{longtable}[c]{|c| c | c | c | c ||  c| c | c | c | c | }
\caption[Assignment of backbone N-H pairs.]{Chemical shift assignment of CypA backbone $^{15}$N-$^1$H pairs, for the unbound protein as well as the CypA-CsA complex. Chemical shifts are given in ppm. Blank cells represent unassigned signals.}\\
\endfirsthead
\hline
\multirow{2}{*}{Residue}& \multicolumn{2}{|c|}{CypA only} & \multicolumn{2}{|c||}{CypA-CsA} & \multirow{2}{*}{Residue}& \multicolumn{2}{|c|}{CypA only} & \multicolumn{2}{|c|}{CypA-CsA} \\
\cline{2-5} \cline{7-10}
& $^1$H	&	$^{15}$N  & $^1$H	& $^{15}$N && $^1$H & $^{15}$N  & $^1$H & $^{15}$N 	\\
\hline
\hline
\endhead
\hline
\endfoot
\hline
\multirow{2}{*}{Residue}& \multicolumn{2}{|c|}{CypA only} & \multicolumn{2}{|c||}{CypA-CsA} & \multirow{2}{*}{Residue}& \multicolumn{2}{|c|}{CypA only} & \multicolumn{2}{|c|}{CypA-CsA} \\
\cline{2-5} \cline{7-10}
& $^1$H	&	$^{15}$N  & $^1$H	& $^{15}$N && $^1$H & $^{15}$N  & $^1$H & $^{15}$N 	\\
\hline
\hline
Thr5	&	8.78	&	115.02	&	8.76	&	115.03	&	Glu86	&	9.45	&	131.63	&	9.51	&	131.82	\\
Val6	&	8.74	&	120.4	&	8.71	&	120.65	&	Asn87	&	7.05	&	106.84	&	7.04	&	106.78	\\
Phe7	&	8.95	&	119.23	&	8.96	&	119.34	&	Phe88	&	8.33	&	113.02	&	8.32	&	112.9	\\
Phe8	&	9.55	&	117.03	&	9.56	&	117.01	&	Ile89	&	8.3	&	119.89	&	8.29	&	119.85	\\
Asp9	&	9.27	&	124.24	&	9.27	&	124.22	&	Leu90	&	7.74	&	117.28	&	7.75	&	117.2	\\
Ile10	&	9.03	&	124.31	&	9.04	&	124.27	&	Lys91	&	8.07	&	119.05	&	8.07	&	119	\\
Ala11	&	9.62	&	132.57	&	9.63	&	132.48	&	His92	&	10.7	&	122.57	&	10.58	&	122.33	\\
Val12	&	8.93	&	118.47	&	8.94	&	118.31	&	Thr93	&	7.26	&	110.48	&	7.25	&	110.42	\\
Asp13	&	9.85	&	131.09	&	9.86	&	131	&	Gly94	&	7.48	&	107.31	&	7.53	&	107.37	\\
Gly14	&	8.55	&	101.75	&	8.56	&	101.63	&	Pro95	&	N/A	&	N/A	&	N/A	&	N/A	\\
Glu15	&	8.04	&	123.34	&	8.03	&	123.33	&	Gly96	&	9.27	&	110.31	&	9.25	&	110.7	\\
Pro16	&	N/A	&	N/A	&	N/A	&	N/A	&	Ile97	&	6.76	&	121.49	&	6.77	&	121.31	\\
Leu17	&	9.2	&	126.13	&	9.17	&	125.96	&	Leu98	&	7.87	&	128.89	&	7.81	&	129.95	\\
Gly18	&	7.24	&	102.52	&	7.23	&	102.33	&	Ser99	&	8.28	&	118.81	&	7.94	&	119.4	\\
Arg19	&	8.34	&	121.29	&	8.35	&	121.28	&	Met100	&		&		&	8.28	&	123.27	\\
Val20	&	9.37	&	126.94	&	9.38	&	126.92	&	Ala101	&	7.99	&	125.97	&	8.22	&	126.95	\\
Ser21	&	8.76	&	120.36	&	8.77	&	120.39	&	Asn102	&	8.11	&	113.4	&	7.68	&	113.59	\\
Phe22	&	9.51	&	119.11	&	9.5	&	119.02	&	Ala103	&	8.77	&	123.37	&	9.18	&	121.03	\\
Glu23	&	8.74	&	123.23	&	8.7	&	123.15	&	Gly104	&	8.18	&	109.38	&	8.24	&	107.71	\\
Leu24	&	8.17	&	122.5	&	8.14	&	122.45	&	Pro105	&	N/A	&	N/A	&	N/A	&	N/A	\\
Phe25	&	8.82	&	124.75	&	8.87	&	124.77	&	Asn106	&	8.86	&	119.13	&	8.87	&	119.05	\\
Ala26	&	8.43	&	129	&	8.46	&	129.05	&	Thr107	&	10.21	&	110.39	&	10.3	&	110.31	\\
Asp27	&	8.97	&	114.09	&	8.99	&	113.99	&	Asn108	&	7.37	&	120.61	&	7.42	&	120.56	\\
Lys28	&	7.53	&	118.08	&	7.53	&	118	&	Gly109	&	9.15	&	110.83	&	9.14	&	110.91	\\
Val29	&	8.36	&	114.64	&	8.36	&	114.86	&	Ser110	&	8.76	&	117.15	&	8.77	&	117.12	\\
Pro30	&	N/A	&	N/A	&	N/A	&	N/A	&	Gln111	&	8.37	&	124.52	&	8.41	&	124.79	\\
Lys31	&	10.66	&	123.84	&	10.7	&	124.02	&	Phe112	&	8.05	&	117.86	&	8.15	&	118.29	\\
Thr32	&	10.28	&	124.06	&	10.27	&	124.08	&	Phe113	&	9.8	&	116.5	&	9.83	&	115.02	\\
Ala33	&	9.28	&	125.67	&	9.23	&	125.52	&	Ile114	&		&		&	9.02	&	118.07	\\
Glu34	&	8.02	&	117.28	&	8.03	&	117.29	&	Cys115	&	9.59	&	125.52	&	9.49	&	124.78	\\
Asn35	&	7.12	&	115.69	&	7.13	&	115.69	&	Thr116	&	8.95	&	115.72	&	9.04	&	115.71	\\
Phe36	&	7.01	&	117.92	&	7.02	&	117.89	&	Ala117	&	7.61	&	122.32	&	7.58	&	122.41	\\
Arg37	&	8.94	&	121.1	&	8.95	&	121.15	&	Lys118	&	8.69	&	119.87	&		&		\\
Ala38	&	8.69	&	119.16	&	8.62	&	118.81	&	Thr119	&	7.61	&	120.21	&	7.37	&	118.3	\\
Leu39	&	8.17	&	120.78	&	8.2	&	120.86	&	Glu120	&	9.09	&	124.63	&	9.06	&	124.65	\\
Ser40	&	7.89	&	119.28	&	7.89	&	119.3	&	Trp121	&	7.25	&	117.92	&	7.3	&	118.3	\\
Thr41	&	7.97	&	108.49	&	7.98	&	108.56	&	Leu122	&	7.01	&	120.05	&	7.05	&	120.72	\\
Gly42	&	7.57	&	108.4	&	7.57	&	108.24	&	Asp123	&	7.59	&	122.24	&	7.56	&	122.37	\\
Glu43	&	8.01	&	118.7	&	8.01	&	118.7	&	Gly124	&	9.53	&	111.33	&	9.57	&	111.14	\\
Lys44	&	9.1	&	118.55	&	9.11	&	118.38	&	Lys125	&	7.74	&	115.68	&	7.57	&	114.75	\\
Gly45	&	7.93	&	105.59	&	7.93	&	105.48	&	His126	&	7.61	&	120.21	&		&		\\
Phe46	&	6.41	&	113.75	&	6.41	&	113.59	&	Val127	&		&		&	8.24	&	124.74	\\
Gly47	&	7.75	&	104.7	&	7.72	&	104.61	&	Val128	&	9.47	&	133.07	&	9.38	&	132.94	\\
Tyr48	&	6.88	&	113.79	&	6.86	&	113.89	&	Phe129	&	8.1	&	117.87	&	8.11	&	117.89	\\
Lys49	&	8.47	&	124.92	&	8.48	&	124.84	&	Gly130	&	7.36	&	110.79	&	7.36	&	110.73	\\
Gly50	&	9.48	&	117.9	&	9.52	&	117.97	&	Lys131	&	8.34	&	115.36	&	8.3	&	115.06	\\
Ser51	&	8.38	&	116.5	&	8.4	&	116.58	&	Val132	&	9.03	&	124.26	&		&		\\
Cys52	&	10.01	&	115.3	&	10.1	&	115.31	&	Lys133	&	9.47	&	131.87	&	9.44	&	131.73	\\
Phe53	&	8.69	&	123.05	&	8.66	&	122.99	&	Glu134	&	7.53	&	118.49	&	7.53	&	118.5	\\
His54	&	7.56	&	119.88	&	7.52	&	119.7	&	Gly135	&		&		&		&		\\
Arg55	&	7.04	&	123.14	&	6.87	&	124.3	&	Met136	&	8.85	&	122.66	&	8.86	&	122.66	\\
Ile56	&	9.19	&	126.48	&	9.13	&	127.27	&	Asn137	&	9.05	&	114.4	&	8.91	&	114.42	\\
Ile57	&	8.54	&	123.02	&	8.49	&	127.36	&	Ile138	&	7.64	&	124.28	&	7.64	&	124.24	\\
Pro58	&	N/A	&	N/A	&	N/A	&	N/A	&	Val139	&	7.24	&	121.97	&	7.26	&	122.16	\\
Gly59	&	9.73	&	114.24	&	9.85	&	114.11	&	Glu140	&	8.26	&	117.69	&	8.31	&	117.82	\\
Phe60	&	8.17	&	119.37	&	8.1	&	119.19	&	Ala141	&	7.49	&	121.13	&	7.49	&	121.04	\\
Met61	&	8.08	&	111.2	&	7.86	&	110.41	&	Met142	&	8.29	&	117.3	&	8.29	&	117.41	\\
Cys62	&	8.47	&	114.88	&		&		&	Glu143	&	7.82	&	116.36	&	7.89	&	116.6	\\
Gln63	&	8.72	&	127.99	&	9.33	&	127.5	&	Arg144	&	7.02	&	114.61	&	7.02	&	114.63	\\
Gly64	&	7.36	&	110.79	&	7.48	&	110.77	&	Phe145	&	7.61	&	115.42	&	7.62	&	115.25	\\
Gly65	&	9.36	&	106	&	9.1	&	105.22	&	Gly146	&	7.5	&	104.75	&	7.56	&	104.82	\\
Asp66	&	9.96	&	124.06	&	10.01	&	124.4	&	Ser147	&	8.2	&	110.11	&	8.19	&	109.89	\\
Phe67	&	6.62	&	116.04	&	6.69	&	116	&	Arg148	&		&		&		&		\\
Thr68	&	7.28	&	109.02	&	7.25	&	108.73	&	Asn149	&		&		&		&		\\
Arg69	&	8.65	&	122.09	&	8.68	&	122.05	&	Gly150	&	8.04	&	110.22	&	8.06	&	110.65	\\
His70	&		&		&	6.53	&	111.11	&	Lys151	&	7.53	&	119.89	&	7.47	&	119.8	\\
Asn71	&	7.49	&	112.49	&	7.44	&	112.24	&	Thr152	&	8.85	&	116.65	&	8.86	&	116.81	\\
Gly72	&	9.66	&	110.6	&	9.67	&	111.25	&	Ser153	&	9.41	&	117.05	&	9.46	&	117.28	\\
Thr73	&	7.93	&	112.2	&	7.86	&	109.83	&	Lys154	&	7.53	&	119.37	&	7.53	&	119.27	\\
Gly74	&	8.71	&	114.04	&	8.8	&	114.03	&	Lys155	&	8.78	&	121.73	&	8.82	&	121.79	\\
Gly75	&	8.11	&	109.15	&	8.06	&	108.9	&	Ile156	&	9.61	&	134.42	&	9.62	&	134.33	\\
Lys76	&	6.97	&	115.71	&	6.93	&	115.5	&	Thr157	&	9.24	&	117.07	&	9.22	&	116.94	\\
Ser77	&	7.79	&	114.41	&	7.75	&	114.44	&	Ile158	&	8.58	&	121.7	&	8.59	&	121.66	\\
Ile78	&	8.55	&	111.24	&	8.55	&	111.1	&	Ala159	&	8.87	&	132.53	&	8.87	&	132.53	\\
Tyr79	&	8.03	&	120.83	&	8.06	&	120.86	&	Asp160	&	8.06	&	111.55	&	8.06	&	111.51	\\
Gly80	&	7.1	&	106.64	&	7.1	&	106.46	&	Cys161	&	8.58	&	116.13	&	8.59	&	116.09	\\
Glu81	&		&		&		&		&	Gly162	&	6.84	&	104.18	&	6.84	&	104.04	\\
Lys82	&	7.86	&	112.72	&	7.83	&	112.28	&	Gln163	&	9.05	&	121.01	&	9.08	&	121.11	\\
Phe83	&	9.17	&	116.64	&	9.2	&	116.47	&	Leu164	&	8.59	&	126.15	&	8.59	&	125.92	\\
Glu84	&	9.23	&	119.54	&	9.28	&	119.46	&	Glu165	&	8.13	&	126.36	&	8.12	&	126.08	\\
Asp85	&	8.59	&	118.9	&	8.62	&	118.81	&		&		&		&		&		
\label{tab:assign}
\end{longtable}
%%%%%%%%%%%%%%%%%%%%%%%%%%%%%%%%%%%%%%%%%%%%%%%%%%

\fi\end{document}